\documentstyle[prl,aps,epsf]{revtex}

\def\nuc#1#2{\relax\ifmmode{{}^{#1}{\relax\text{#2}}}\else{${}^{#1}$#2}\fi}
\def\nuc#1#2{${}^{#1}$#2}

\begin{document}
\draft

\title{Nucleation and Interfacial Coupling between Pure and Dirty Superfluid Phases of $^{3}$He}

\author{G. Gervais, K. Yawata, N. Mulders$^\dag$, and W.P. Halperin}

\address{Department of Physics and Astronomy, Northwestern University, Evanston, Illinois, 60208 USA\\
$^{\dag}$Department of Physics and Astronomy, University of Delaware, Newark, Delaware, 19716 USA\\
}

\date{\today}

\twocolumn[\hsize\textwidth\columnwidth\hsize\csname@twocolumnfalse\endcsname

\maketitle
\begin{abstract}
The nucleation of the first order phase transition of superfluid $^{3}$He-B from superfluid $^{3}$He-A is quite
remarkable since it requires a seed of the order of a micron.  We have studied this nucleation for
$^{3}$He confined to a very dilute silica aerogel.  This dirty superfluid behaves in a manner similar to previous reports
for the pure superfluid. But we have discovered a novel magnetically driven nucleation switch acting on the pure
superfluid B-phase. Lastly, we find the surprising result that the proximity effect between the pure and dirty
superfluids at their interface is insufficient to nucleate the B-phase in either superfluid.
 
\end{abstract}
]
\narrowtext 

The A-phase of superfluid $^{3}$He can be extensively supercooled, far into the region at low temperatures
where the B-phase is thermodynamically stable\cite{Schiffer92,Schiffer95}. 
Leggett\cite{Leggett84,Leggett89} pointed out that the very small difference in free energy between these phases makes
homogeneous nucleation of this first order phase transition so improbable as to be unrealizable within the age of the
universe.  If a seed of the B-phase is to grow it must surpass a critical size of a few microns, so large as to be
inaccessible by thermal fluctuations.  Thus the experimental fact that the B-phase nucleates at all from the A-phase
indicates the existence of some heterogeneous mechanism.  Quite a number of experiments have shown that ionizing
radiation\cite{Schiffer92,Schiffer95}, vibrations\cite{Okeefe,Pickett00}, or rough surfaces\cite{Okeefe,Hakonen} can be,
under various restrictive circumstances, active sites for this nucleation; but how the process proceeds remains a major
puzzle and is actively debated\cite{Bunkov98}.

The discovery of a class of $^{3}$He superfluids\cite{Porto95,Spra95} constrained by silica aerogel raises new
questions concerning B-phase nucleation.   Does the  metastability of the A-phase for these superfluids follow the same
pattern as for the pure case? Is there sufficient coupling between the two superfluids across their interface
for nucleating the B-phase? With an
appropriate choice of experimental conditions, temperature, pressure, and magnetic field, we can arrange that the interface
separates two superfluids with the same, or different, order parameter symmetry.  And if they are the same,  we might
expect the proximity effect, well known in superconductivity, to provide a nucleation source.  By this we mean
that the B-phase from one side of the interface should readily nucleate the B-phase in the other.   We report here that
indeed the superfluid constrained in aerogel exhibits supercooling and metastability similar to that of pure superfluid
$^{3}$He in the absence of aerogel; however, contrary to our expectation, there is no
evidence of nucleation from the proximity effect.  In addition, we have discovered a nucleation source for the pure
B-superfluid that is extremely efficient and can be switched off by applying a magnetic field.

The new `dirty'  superfluids are characterized by quasiparticle scattering from a silica aerogel-matrix that reduces their
transition temperature and suppresses the amplitude of the order parameter\cite{Spra95,Spra96}.  The aerogel structure can
be varied through choice of porosity, which in this work is 98\%.  It is formed by silica strands about \protect{$30$ \AA}
in diameter with an average inter-strand spacing of
$300$ \AA.  The pure superfluid coherence length, $\xi$, varies from \protect{$880$ \AA}  at low pressure (zero bar) to
\protect{$180$ \AA} at melting pressure (34 bar) and is much larger than the aerogel microstructure, yet much less than the
scattering mean free path $\sim 2500$ \AA\cite{Nomura00}.  This satisfies conditions for the existence of a superfluid, albeit
one with a reduced order parameter.  The phase diagrams of the pure\cite{Hahn95} and dirty\cite{Gervais01} systems have some
similarity as shown in Fig.1 and 2.  
The magnetic field independent line marks the normal to superfluid A-phase boundary and the transition temperature from
A to B-phases is quadratically suppressed by field. The first 
observation of an AB-transition in the dirty superfluid was reported by Barker \textit{et al.}\cite{Barker00}. We
assume that the order parameter symmetry of the dirty phases corresponds to that of the pure
$^{3}$He superfluids, \textit{i.e.} the dirty A-phase is an equal-spin, p-wave pairing state called the axial state,
exhibiting broken rotation symmetry separately in spin and orbital spaces, and that the B-phase is the isotropic p-wave
pairing state that breaks relative spin-orbit symmetry.

First-order phase transitions, such as the the liquid-solid transition, characteristically exhibit supercooling. 
Nucleation of the stable phase can be understood in terms of a trade-off in surface and bulk Gibbs free energy.  The
nucleation theory developed by Gibbs and discussed by Landau and Lifshitz, and later extended by Cahn and Hilliard
considers a small embryo of the stable phase in the metastable medium.  If the embryo is sufficiently large and
exceeds a critical radius $R_{c}$,  it will expand over the whole volume. This critical radius,
$R_{c} = 2\sigma_{AB}/\Delta G_{AB}$ is given by twice the ratio of the surface tension,
$\sigma_{AB}$, between the two phases,  A and B, and the bulk Gibbs free energy difference, $\Delta G_{AB}$. There are
several examples of quantum fluids and solids at low temperatures, where $\Delta G$ is very small, forcing the critical
radius to be rather large, leading to extensive supercooling. This is true for the liquid-solid transition of 
$^{4}$He\cite{Balibar00} or the  transition between superfluid $^{3}$He A and B-phases\cite{Leggett89}. 
\begin{figure}[t]
  \begin{center}
   \leavevmode
    \hbox{\epsfxsize=1.\columnwidth{\epsfbox{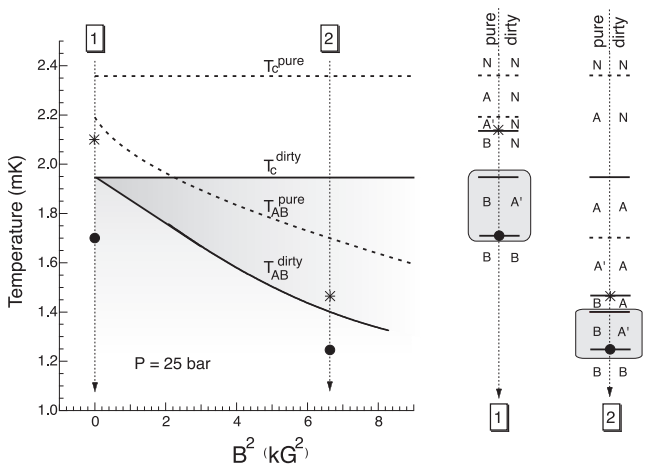}}}
\end{center}
\caption{The phase diagrams of the superfluid phases of $^{3}$He at a pressure of 25 bar in a magnetic field.  The
phase diagrams for pure\protect\cite{Hahn95} and dirty\protect\cite{Gervais01} $^{3}$He are given by dashed and solid lines
with the dirty superfluid region shaded. The darker shading identifies the dirty A-phase; the
lighter shading is the dirty B-phase. Two constant field experiments are shown as dotted vertical
lines with arrows and, at the right of the figure on the same temperature scale, we show the corresponding phases observed
(N is the normal  Fermi liquid, A and B are the superfluid phases, and A' is the metastable A-phase). Primary
nucleation of the pure (star) and  the dirty (filled
circle) AB-transitions appear at, or below, their respective equilibrium curves. The shaded rectangular areas at the right
emphasize the coexistence of A' and B across the interface.}
\label{phase25}
\end{figure}

Our acoustic technique\cite{Gervais01} measures the transverse acoustic impedance
at $8.691$ MHz of a gold-plated quartz transducer. Two transducers separated by a distance of 270 $\mu$m \cite{GervaisJLTP}
define a
$9.5$mm diameter disk of silica aerogel grown \textit{in situ}. The interface region between pure and
dirty $^{3}$He is a circular band of 270 $\mu$m width at the perimeter of the transducers. We observed abrupt changes in
impedance at all of the known phase transitions in each of pure and dirty superfluids \cite{Gervais01}. The technique is very
precise and is sensitive over almost all of the phase diagram.  We have used it to map  the
phase diagram of  the superfluid phases inside the aerogel sample\cite{Gervais01}, together with monitoring the pure
superfluid transitions.  Our thermometers include the magnetization of a diluted cerious magnesium nitrate salt measured with
a SQUID, complemented by a melting curve thermometer. Overall our precision in the measurement of temperature 
is $\sim 2\mu$K with an accuracy on cooling of $\sim20$ $\mu$K.

\begin{figure}[t]
  \begin{center}
   \leavevmode
    \hbox{\epsfxsize=1.0\columnwidth{\epsfbox{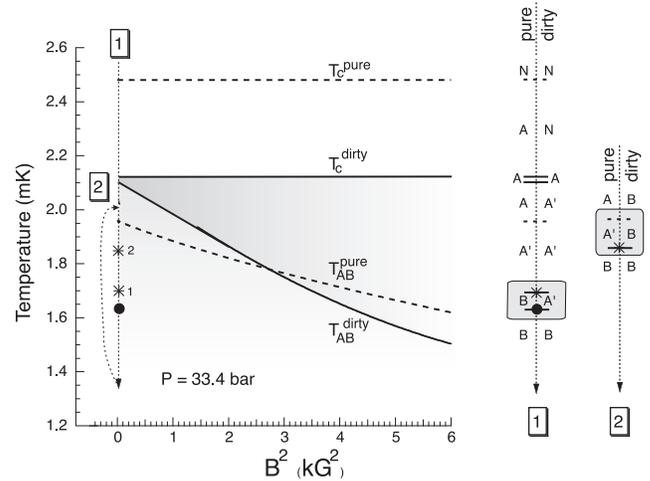}}}
  \end{center}

\caption{The phase diagrams of the superfluid phases of $^{3}$He at a pressure of 33.4 bar in a magnetic
field\protect\cite{Hahn95,Gervais01}. The notation is defined in Fig.\protect\ref{phase25}. The curved
dotted-path labeled 2 is an experiment to study secondary nucleation in pure $^{3}$He.  The temperature was raised above
$T_{AB}^{pure}$, yet kept below $T_{c}^{pure}$. The stars labeled 1 and 2 denote the temperatures for primary and secondary
nucleation of the pure B-phase. }

\label{diagram}
\end{figure}

We have performed both primary and secondary nucleation experiments.  Primary nucleation of the B-phase, from the supercooled
A-phase, occurs on cooling provided that there has been no prior history of B-phase. Secondary
nucleation occurs on supercooling of the A-phase after a primary nucleation, but without having warmed to the normal state.
Representative nucleation experiments are outlined in Fig.1 at 25 bar and Fig.2 at 33.4 bar, as
vertical dotted lines superposed on the phase diagrams for pure
$^{3}$He (dashed lines) \cite{Hahn95} and dirty $^{3}$He (solid lines) \cite{Gervais01}. The
shading corresponds to the region of the dirty superfluid, where the darker area is the A-phase in equilibrium.  
Supercooling of the A to B-transition is observed for both pure (star) and dirty superfluids (solid circle) with all data for
primary nucleation collected in Fig.3. The path for a secondary nucleation experiment in the pure superfluid is sketched in
Fig.\ref{diagram} with the results in Fig.4.

We have found that primary nucleation in the pure and dirty superfluid (Fig.3) are similar 
with one exception which we discuss later: for pure $^{3}$He at 25 bar in fields less than $2$ kG nucleation can be
efficiently induced by a novel nucleation source. Otherwise, supercooling for both pure and dirty superfluids is
similar in magnitude, field independent below $2$ kG, somewhat stronger at higher pressures (open circles), but more
stochastic for pure
$^{3}$He. Earlier reports of supercooling in pure
$^{3}$He have shown sensitivity to outside influences such as radiation\cite{Schiffer95}, but the most efficient nucleation,
seems to take place at rough surfaces such as in the silver sintered powder of the heat exchanger required to cool samples to
low temperatures\cite{Okeefe,Hakonen}.  With the exception noted above, we find supercooling for pure
$^{3}$He to be consistent in magnitude with that previously discussed for rough walls. The similarities between
the pure and dirty superfluids are quite surprising
since they are in very different environments and presumably have access to different nucleation sites. 
Furthermore, as we discuss later, the critical radius for nucleation is significantly different 
for the two superfluids.

  \begin{figure}[t]
  \begin{center}
   \leavevmode
    \hbox{\epsfxsize=0.85\columnwidth{\epsfbox{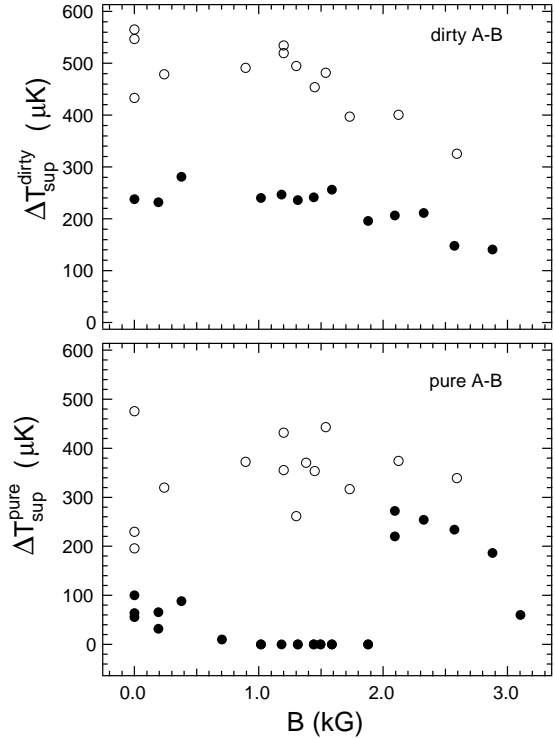}}}
  \end{center}
  \caption{Supercooling for primary nucleation. $\Delta T_{sup}\equiv T_{AB}-T$ where T is the temperature at which
nucleation occurs. The dirty AB-transition (upper) and pure AB-transition (lower) are shown at 33.4 bar (empty circles)  and
25 bar (filled circles).}
\label{primary}
\end{figure} 

The one exception to this picture is the 25 bar data where we found extremely efficient nucleation of the pure
AB-transition for magnetic fields less than $2$ kG. This nucleation
source switches off when a field larger than 2 kG is applied and exhibits hysteresis; on reducing the field to 
$\sim$1 kG from $\sim$2.5 kG the nucleation source remains inactive. The data at 33.4 bar, Fig.\ref{primary} lower
panel, were taken with the switch deactivated in this way. They show supercooling to be large, $\sim350
\mu$K. The strong nucleation source is restored (switched on) following a warmup of the experimental cell to
room temperature. These observations, together with  the hysteretic behavior
in field, suggest that the nucleation source has a magnetic origin which we have not yet identified.

\begin{figure}[t]
  \begin{center}
   \leavevmode
    \hbox{\epsfxsize=0.9\columnwidth{\epsfbox{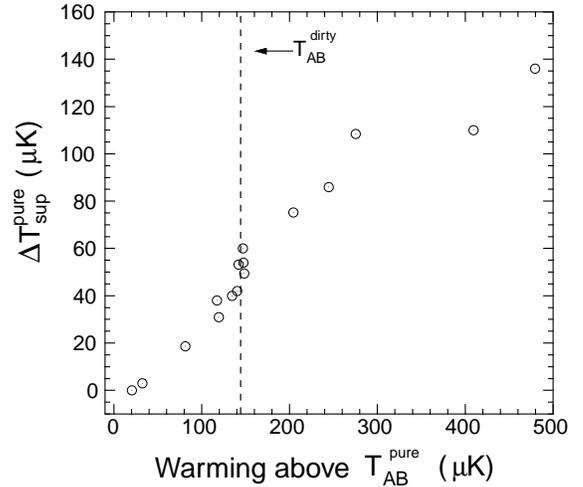}}}
  \end{center}
  \caption{Supercooling memory effect for secondary nucleation of the pure AB-transition at 33.4 bar and zero applied field. 
Supercooling of the AB-transition depends on how much the sample was warmed above the equilibrium $T_{AB}^{pure}$. The
vertical dashed line indicates $T_{AB}^{dirty}$. To the left of this line, the $^{3}$He in the aerogel was in the  B-phase
when the pure B-phase was nucleated as for the experiment sketched in Fig.2; on the right it was in A'.}
  \label{secondary}
\end{figure}

The effectiveness of heterogeneous nucleation depends on the critical
radius at the temperature where nucleation takes place.  To estimate this for the dirty superfluid we note that Osheroff and
Cross\cite{Osheroff77} have established a reasonable theoretical understanding of the surface tension between A and
B-phases, summarized by Leggett and Yip\cite{Leggett89},
$\sigma_{AB} \sim F_{s}\xi$. The coherence length  $\xi$ varies as $1/T_{c}$ and the condensation energy 
$F_{s}$ can be calculated from the suppression of the order parameter in the dirty superfluid \cite{foot1}. 
We find $\sigma_{AB}^{dirty}\sim 0.25 \sigma_{AB}^{pure}$ near melting pressure. The free energy
difference between the phases, $G_{AB}$, can be determined from their susceptibility difference and the field dependence of
the AB-transition\cite{Osheroff77}. The susceptibility difference in the dirty superfluid has been
measured\cite{Spra96,Barker00} at 18 and 32 bar, and is approximately a factor of two weaker than for the pure
superfluid. We have measured the field dependence of the dirty AB-transition at 25 and 33.4 bar, Fig.1 and 2.
Consequently, the estimated critical radius for nucleation of the dirty B-phase is larger than that for the pure
superfluid,
$R_{c}^{dirty}(t)\sim 5R_{c}^{pure}(t)$, at the same reduced temperature
$t\equiv T/T_{AB}$ and near melting pressure where $R_{c}^{pure} \sim$ 1 micron. With a critical radius for the dirty
superfluid somewhat larger than for pure
$^{3}$He we might expect larger supercooling, and the data have this trend. At 25 bar, for magnetic fields between
0.5 and 2 kG, supercooling in pure
$^{3}$He is quenched implying that a different mechanism is active, one that generates a huge nucleation seed. Since
supercooling was  at most
$\sim 10$ $\mu$K, this seed must have been larger than  $\sim$ 70 microns \cite{Osheroff77}.

The primary nucleation experiments we have discussed allow us to study the juxtaposition of a pure B-phase and a metastable
dirty A-phase. We can also reverse the roles of pure and dirty superfluids but this requires $T_{AB}^{dirty}>
T_{AB}^{pure}$ which is possible at 33.4 bar and low field (path 2 in Fig. 2). In this case the nucleation
will be secondary. Secondary nucleation in pure $^{3}$He was first observed by Osheroff \textit{et al.}
\cite{OsheroffTh}. Remnants of the B-phase persist above the AB-transition temperature giving rise to a memory effect for
nucleation on subsequent cooling. We also observe a memory effect in the pure system but in
contrast there is none in the dirty superfluid. It is quite striking that
the memory effect evident in the pure superfluid data of Fig.4 is unperturbed on crossing the AB-phase boundary of
the dirty superfluid.  This shows that the presence or absence of a B-superfluid in the aerogel has no effect on
nucleation  of the B-phase in pure
$^{3}$He.

Consider the interface between the two superfluids. The rectangular shaded regions at the right of
Fig.1 and 2 correspond to a metastable A-phase, which for convenience we call an A' phase,  in
contact across the interface with a B-phase.  Either pure or dirty superfluids can be in this configuration.  These
experiments show that the B-phase from one superfluid does not act as a nucleation source for the B-phase in the other.
The Gibbs free energy difference between phases on either side of the interface (separating the pure and dirty
superfluids) is relatively large  and can be estimated from the suppression of the gap in the dirty superfluid
\cite{foot1}. If we assume that the surface tension at the interface is of similar magnitude as that of the AB-phase
boundary in the pure and dirty superfluids, then  we calculate that the effective critical radius for penetration of
one phase through the interface is, $R_{c}^{I}(t)\sim R_{c}^{pure}(t)/80$ $\sim$125 \AA. This is so small that we
expect that the B-phase on one side should be an effective nucleation source for the B-phase on the other.  From the
experiments we report here there is no evidence for such nucleation.  Is it possible that the dirty superfluid is
sufficiently  inhomogeneous that a suitably large homogeneous seed cannot be generated? Imry and Ma \cite{Imry78}
found that an order parameter is unstable to any amount of disorder from a random field. So we speculate that the
absence of  nucleation by the proximity effect may be related to decoherence of the order parameter in the dirty
superfluid on the length scale of the critical radius. Another possibility is that the orbital symmetry of the dirty
superfluid is not the isotropic state, as we have assumed.

We have studied nucleation of the B-phase of superfluid $^{3}$He and compared
it to a dirty superfluid formed in an aerogel matrix. The supercooling behavior is similar except 
at low field where a highly efficient nucleation source can be magnetically activated for the pure superfluid
AB-transition. Furthermore, we find that there is no nucleation provided by proximity between pure and dirty
superfluids at their interface.  We suggest an explanation in terms of decoherence of the dirty superfluid order
parameter on the scale of the critical radius.

We acknowledge helpful discussions with J.A. Sauls.  One of the authors (G.G) acknowledges
support from FCAR (Qu\'ebec). This work was supported by the NSF, grant no. DMR-0072350.


\begin{thebibliography}{70}
\bibitem{foot1} We take $F_{s}$ to scale with the square of the energy gap, and  $\Delta_{dirty}^2\simeq 0.4\Delta
_{pure}^2$ \protect\cite{Spra96}.
\end{thebibliography}
\end{document}